\documentclass{aa} 
\usepackage{graphicx}
\usepackage{txfonts}
\usepackage[authoryear]{natbib}
\usepackage[rightcaption]{sidecap}

\begin{document}
 
\title{Evidence for Partial Taylor Relaxation from Changes in Magnetic Geometry and Energy during a Solar Flare}

\author{Sophie A. Murray \and D. Shaun Bloomfield \and Peter T. Gallagher}

\institute{Astrophysics Research Group, School of Physics, Trinity College Dublin, Dublin 2, Ireland.\\
            \email{somurray@tcd.ie}}

\authorrunning{Murray, S. A., et al}
\titlerunning{Sunspot Magnetic Geometry and Energy Changes during a Solar Flare.}
              
\abstract
{Solar flares are powered by energy stored in the coronal magnetic field, a portion of which is released when the field reconfigures into a lower energy state. Investigation of sunspot magnetic field topology during flare activity is useful to improve our understanding of flaring processes.}
{Here we investigate the deviation of the non-linear field configuration from that of the linear and potential configurations, and study the free energy available leading up to and after a flare.}
{The evolution of the magnetic field in NOAA region 10953 was examined using data from $Hinode$/SOT-SP, over a period of 12 hours leading up to and after a GOES B1.0 flare. Previous work on this region found pre- and post-flare changes in photospheric vector magnetic field parameters of flux elements outside the primary sunspot. 3D geometry was thus investigated using potential, linear force-free, and non-linear force-free field extrapolations in order to fully understand the evolution of the field lines.}
{Traced field line geometrical and footpoint orientation differences show that the field does not completely relax to a fully potential or linear force-free state after the flare. Magnetic and free magnetic energies increase significantly $\sim$~6.5--2.5~hours before the flare by $\sim 10^{31}$~erg. After the flare, the non-linear force-free magnetic energy and free magnetic energies decrease but do not return to pre-flare `quiet' values.}
{The post-flare non-linear force-free field configuration is closer (but not equal) to that of the linear force-free field configuration than a potential one. However, the small degree of similarity suggests that partial Taylor relaxation has occurred over a time scale of $\sim$ 3--4 hours.}

\maketitle

\section{Introduction}
\label{introduction}

Solar flares occur when energy stored in active region magnetic fields is suddenly released. The stored magnetic energy is converted to kinetic energy of particles, mass motions, and radiation emitted across the entire electromagnetic spectrum \citep[see][]{fletcher11}. Energies of large flares can be as high as $\sim 10^{32}$~ergs, over short time scales of tens of minutes. The field configuration of active regions is believed to be linked to the likelihood of flaring; flare triggers are often associated with flux emergence and increased shear and twist in the field \citep[see review by][and references therein]{priest02}. However, the processes involved in magnetic energy storage and release within active regions are still not fully understood. Energy release is expected to occur in the corona, but magnetic field observations are usually based in the photosphere. In order to investigate the coronal magnetic field, photospheric measurements are used as a starting point for 3D magnetic field extrapolations \citep{gary89}. Although there are inaccuracies involved with the various assumptions that must be made to obtain 3D extrapolations, much progress has been made in recent years with high resolution photospheric magnetic field measurements now available from spacecraft such as $Hinode$ \citep{kosugi07} and \emph{Solar Dynamics Observatory} \citep{pesnell12}. Increasingly accurate 3D magnetic field extrapolations have yielded insight into active region magnetic field topology during periods of flare activity  \citep{regniercanfield06, thalmannwiegelmann08, sun12}.


This paper investigates the evolution of the 3D coronal magnetic field in an active region using three types of extrapolation procedure: potential, linear force free (LFF), and non-linear force free (NLFF). The corona is considered to be generally force free \citep{gold60}, dominated by the relatively stable magnetic field in a low-$\beta$ plasma. Hence, all three types of 3D extrapolation assume the force-free approximation, $j \times B = 0$, where $j$ is the current density and $B$ is the magnetic field \citep{gary01}. This assumption is made under the special conditions of magnetohydrostatic equilibrium. The approximation results in the current being proportional to the magnetic field, with a proportionality factor $\alpha$ termed the force-free field parameter. There are three generalised forms of the force-free relation, 
\begin{equation}
\label{potential}
\nabla \times B = 0,
\end{equation}
\begin{equation}
\label{lff}
\nabla \times B = \alpha B,
\end{equation}
\begin{equation}
\label{nlff}
\nabla \times B = \alpha(x,y,z) B.
\end{equation}
A potential field configuration is defined as one containing no currents, resulting in the case of Eqn.~\ref{potential} where $\alpha=0$. When $\alpha$ is non-zero but constant throughout a given volume the field configuration is referred to as LFF (Eqn.~\ref{lff}). Finally, when $\alpha$ is allowed to vary spatially (i.e., differing from field line to field line, but constant along one field line) the field configuration is referred to as NLFF (Eqn.~\ref{nlff}). This specific case allows for the existence of both potential and non-potential fields within the given volume. 

Potential field extrapolations can be thought of as the first-order approximation to the coronal magnetic topology. They are often used to represent solar global magnetic fields, which are dominated by simple dipolar configurations with few currents \citep{liu08}. In contrast, LFF field extrapolations have been considered sufficient to represent the large-scale, current-carrying coronal magnetic fields present on whole active region size scales \citep{gary89, wheatland99}. However, NLFF field extrapolations intrinsically contain more information on the complex nature of solar magnetic fields and more accurately represent coronal magnetic structures on size scales smaller than whole active regions. For example, \citet{Wiegelmann05} compared field extrapolations from photospheric measurements to observed chromospheric magnetic fields, finding the potential solution provided no agreement with any observed field lines, the LFF solution provided 35\% agreement, while the NLFF solution provided 64\% agreement. It should be noted that there are still inconsistencies between existing NLFF extrapolation methods, particularly regarding the treatment of boundary conditions \citep[see, e.g., the comparison paper of][]{derosa09}.

The underlying principle of force-free coronal fields has previously been extensively studied. Initially, \citet{woltjer58} theoretically examined a magnetic field configuration over the course of magnetic energy release, proposing that force-free fields with constant $\alpha$ (LFF) represent the state of lowest magnetic energy in a closed system. \citet{taylor86} applied this theory to laboratory plasma experiments, suggesting that the total magnetic helicity of a flux system is invariant during the relaxation process to this minimum-energy state. The theory and experimental evidence led to the concept that the free magnetic energy that may be released during field relaxation in a solar active region is the excess energy above the LFF field with the same magnetic helicity \citep{heyvaerts84}. This relaxation process will henceforth be referred to as Taylor relaxation throughout this paper. Several theoretical, numerical, and observational studies have investigated whether Taylor relaxation occurs during solar flares, with some agreeing to its presence \citep[e.g.,][]{nandy03, browning08} and others disagreeing \citep[e.g.,][]{amari00, bleybel02}.

Investigating the magnetic and free magnetic energy in active regions is essential to study the physical processes occurring during solar flares. Previously, decreasing magnetic energy has been reported after flares \citep[e.g.,][]{regniercanfield06}. Also, energy budgets for combined flare-CME events were studied in detail by \citet{emslie05}, finding $\sim$20--30\% of the available free magnetic energy was required to power the events.


In this paper, we examine active region magnetic energy evolution before and after a flare (on much shorter timescales than previously studied), in order to gain a better understanding of the processes involved in energy release. This paper compares the results of three different forms of 3D magnetic field extrapolation by examining the evolution of an active region. In Sect.~\ref{obs} we briefly discuss the observations and analysis techniques employed. Section~\ref{results} presents the main results, in particular geometrical differences in traced field-line solutions (Sect.~\ref{loops}), orientation differences in field-line footpoint solutions (Sect.~\ref{res1}), and the evolution of magnetic energy and free magnetic energy (Sect.~\ref{energy_res}). Finally, our conclusions and directions for future work are outlined in Sect.~\ref{concl}.

\section{Observations}
\label{obs}

Active region NOAA 10953\footnote{http://solarmonitor.org/region.php?date=20070426\&region=10953}~was observed by the \emph{Hinode} Solar Optical Telescope \citep[SOT:][]{tsuneta08} as it crossed the solar disk on 2007 April 29. The primary instrument used in this work from the SOT suite is the spectropolarimeter (SP). SOT-SP records the Stokes $I$, $Q$, $U$, and $V$ polarization profiles of the Fe\,\textsc{i} 6301.5~\AA\ and 6302.5~\AA\ lines simultaneously through a $0.16'\arcsec \times 164\arcsec$ slit. The data utilised here correspond to SOT-SP fast mapping mode (i.e., $0.32\arcsec \times 0.32\arcsec$ pixels due to co-adding of adjacent slit positions and 2-pixel rebinning along the slit onboard the spacecraft). Details of the complete analysis procedure applied to the data are contained in \citet{murray11} and references therein, including the atmospheric inversion procedure, 180$^\circ$-azimuth disambiguation, and heliographic coordinate conversion. Four SOT-SP scans are examined from 2007 April 29 covering a period of 12~hours, leading up to and after a \emph{GOES} B1.0 solar flare that peaked at 10:37~UT. Three scans were obtained before the flare (starting at 00:17~UT, 03:30~UT, and 08:00~UT, respectively) and one after the flare (starting at 11:27~UT). Note that each scan takes $32$~minutes to be built up from stepping the slit over the chosen field-of-view (FOV). Supplementary information on the location of the flare brightening is obtained from the Ca\,\textsc{ii}\,H filter of the SOT Broadband Filter Instrument.

\begin{figure*}
\centering
\includegraphics[width=\textwidth]{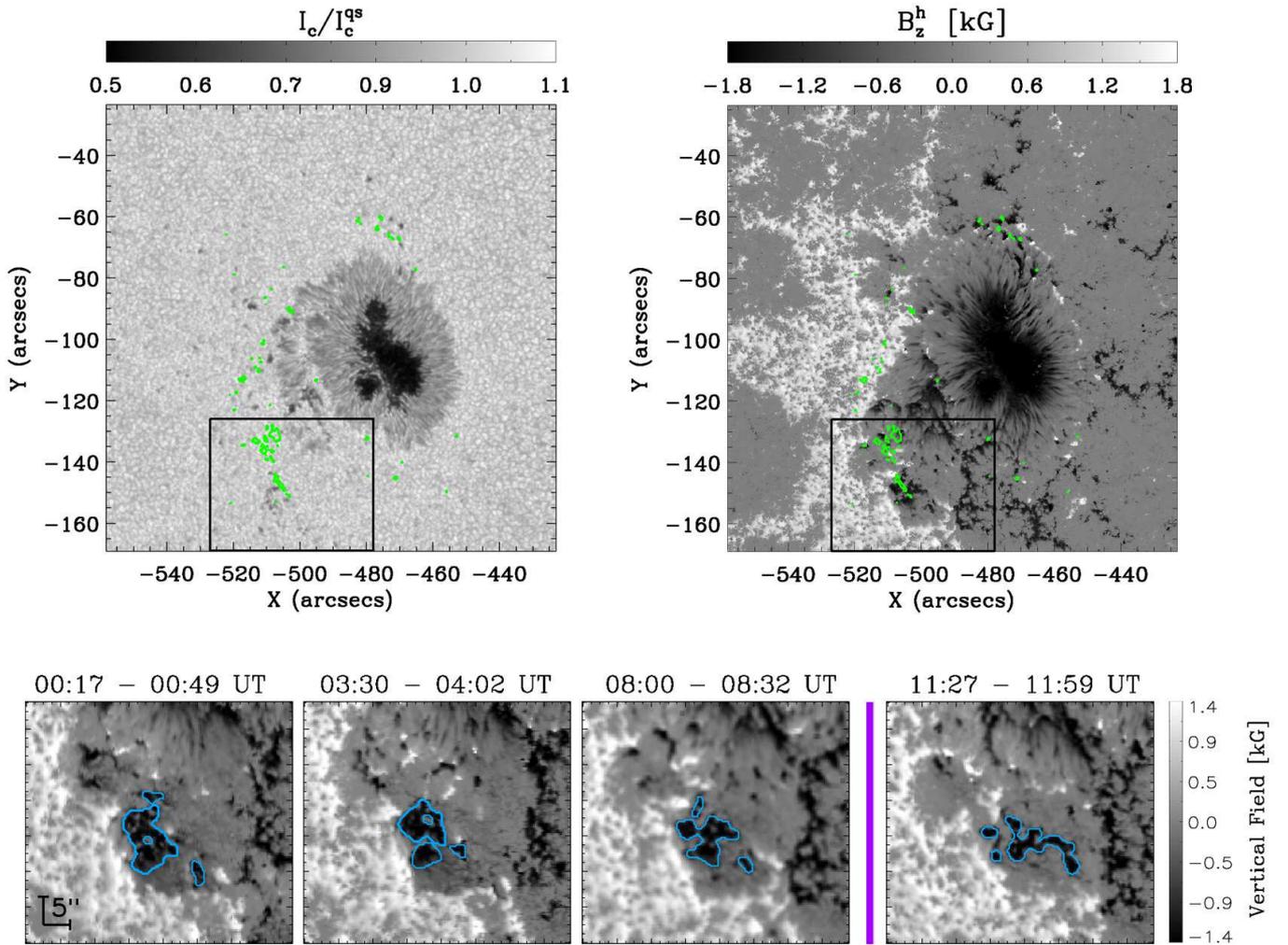}
\caption{Upper row shows the active region pre-flare state in continuum intensity (left) and vertical field strength (right). Green contours overlaid on both images indicate Ca\,\textsc{ii}\,H flare brightening observed at flare peak. The $\sim 50" \times 40"$ sub-region indicated by the black box in the upper row is shown in the lower row as preprocessed vertical field strength, at increasing scan time from left to right. A region of interest, thresholded at $-$750~G, is defined by the blue contours, and the purple line indicates the time of flaring between scans 3 and 4.} 
\label{context}
\end{figure*}

The active region investigated by \citeauthor{murray11} is shown in the top row of Fig.~\ref{context}, which is further investigated in this paper. These panels contain the continuum intensity (upper left) and surface vertical field strength (upper right) of the third scan (i.e., the scan prior to the flare). Previously, \citeauthor{murray11} identified two regions of interest (ROIs; depicted in their Fig.~2) in an area of increased Ca\,\textsc{ii}\,H flare emission. These regions are contained within the box in both upper panels of Fig.~\ref{context} (with green contours showing the Ca\,\textsc{ii}\,H intensity at the flare peak). This zoomed-in box (also shown in the lower row of Fig.~\ref{context}) is used later for a portion of the analysis in Sect.~\ref{energy_res}. The surface magnetic field evolution of this region was previously considered by \citet{murray11}, finding significant changes in vector field parameters in the ROIs during the observation period. Both ROIs showed a change in field inclination towards the vertical before the flare, with the field returning towards the horizontal afterwards. This variation in inclination agrees with inclination changes across the neutral line predicted by \citet{hudson08}.  However, 3D coronal extrapolations are necessary to fully understand the evolution of the field, which is the specific purpose of this paper.

\section{Field Extrapolation Method}

In this work the output vector-magnetic field data analysed by \citeauthor{murray11} is subjected to further analysis. The heliographic vector field information (i.e., $B_{x}^{\mathrm{h}}$, $B_{y}^{\mathrm{h}}$ and $B_{z}^{\mathrm{h}}$) obtained by \citeauthor{murray11} were used as inputs to the three forms of 3D magnetic field extrapolation outlined in Sect.~\ref{introduction}. It is necessary to preprocess photospheric magnetic field data before its use as a boundary condition for 3D extrapolations. The procedure outlined by \citet{wiegelmann06} is used to deal with the inconsistency between the force-free assumption of NLFF models and the non force-free nature of the photospheric magnetic field, while also removing certain noise issues (such as uncertainties in the transverse field components). Note that preprocessing is applied to the data prior to the application of all three extrapolation methods.

The potential and LFF fields were calculated using themethod of \citet{seehafer78}, implemented in the {\tt LINFF} code developed by T. Wiegelmann (2011, private communication). To obtain the potential extrapolation using {\tt LINFF}, $\alpha$ is simply set to zero. For the case of LFF extrapolations, the required values of constant $\alpha$ were calculated by fitting $J_{z}^{\mathrm{h}}$ vs. $B_{z}^{\mathrm{h}}$ from the results of \citeauthor{murray11} \citep[see][]{hahn05}. The values of $\alpha$ obtained for the full FOV of each scan (i.e., $455 \times 455$~pixels$^2$) are $[0.30, 0.29, 0.27, 0.28]~\mathrm{Mm}^{-1}$ for scans 1, 2, 3, and 4, respectively. The values are reasonably similar, showing a slight decrease over time before the flare with a marginal increase afterwards. The chosen form of NLFF extrapolation is the weighted optimization method originally proposed by \citet{wheatland00} and implemented by \citet{wiegelmann04}. This is currently one of the most accurate NLFF procedures available, as demonstrated in the \citet{schrijver06}, \citet{metcalf08}, and \citet{derosa09} method reviews. The NLFF optimization method begins by computing a potential field in the computational volume using the \citet{seehafer78} method. The bottom boundary plane is defined by the input vector magnetogram, with the lateral and top boundary planes described using the potential field results. It is worth noting, this NLFF code directly minimises the force-balance equation, which avoids the explicit computation of $\alpha$.

A useful consistency check for the preprocessing method is the comparison of flux balance, net force, and net torque parameters \citep[see Section 2 of][for more details]{wiegelmann06}. In order to serve as a suitable bottom-boundary condition, vector magnetograms must be approximately flux balanced, and net force and net torque must vanish relative to the force due to magnetic pressure \citep{low85}. The results for the data set used in this paper are promising. For example, the sum of net force and net torque normalised to magnetic pressure before preprocessing was on average $\sim 0.4790$, and after preprocessing this value drops to $\sim 1.6 \times 10^{-3}$. Thus, the preprocessing results in a data set more consistent with the assumption of a force-free coronal magnetic field. The bottom row of Fig.~\ref{context} shows the evolution of the preprocessed surface vertical field strength in the area of increased Ca\,\textsc{ii}\,H intensity (zoom-in region delineated by the box in the upper panels of Fig.~\ref{context}).

\begin{figure}
\centering
\includegraphics[width=\columnwidth]{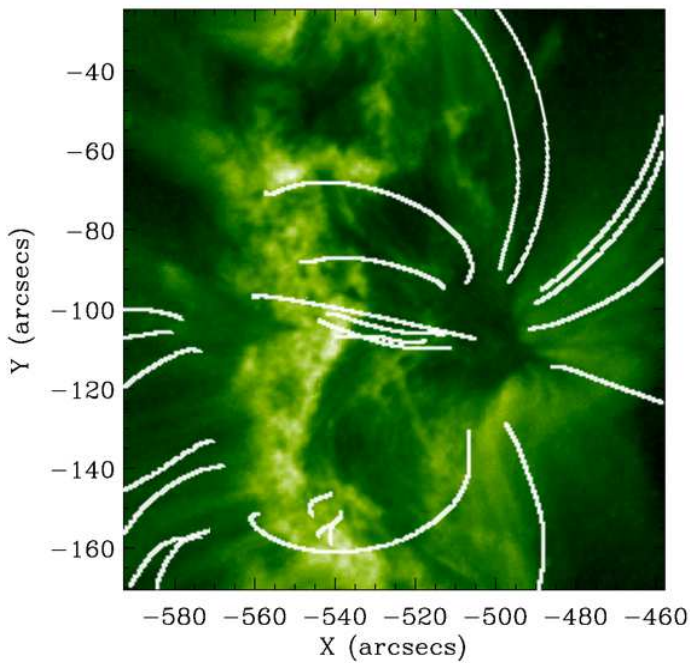}
\caption{TRACE 195~\AA\ log-scale image of NOAA active region 10953 at 03:45~UT on 29 April 2007. The overplotted white lines represent selected field lines from NLFF extrapolation results of scan 2.}
\label{3D:trace}
\end{figure}

As mentioned in \citet{derosa09} and \citet{wiegelmann12}, it is useful to compare the extrapolated field with coronal-loop observations as a consistency check. This quantifies the extent to which the extrapolation correctly reproduces the coronal magnetic field configuration. Unfortunately very few high-resolution coronal observations were available for this particular event. Observations were obtained from the Transition Region and Coronal Explorer \citep[TRACE;][]{handy99} for comparison; a 195~\AA\ image was recorded at 03:45~UT on 29 April 2007 (i.e., during scan 2). The region of flare brightening examined by \citet{murray11} contains only small-scale coronal loops, therefore large-scale loop structures from the full FOV were examined instead. Figure~\ref{3D:trace} shows select NLFF field lines traced over the image. Note that only the NLFF extrapolation is examined, as it is considered to be the most accurate representation of the coronal field (compared to the potential or LFF extrapolations). It is unfortunately difficult to identify coronal loops in this 195~\AA\ wavelength image (171~\AA\ for example would have been more useful), but the extrapolated field lines seem to emanate from the leading sunspot to connect with areas of bright coronal emission. Agreement is particularly clear in the loops traced to the northeast of the image.


A single ROI in the area of flare brightening was selected for further investigation. This was identified by thresholding the preprocessed surface vertical field strength at $-$750~G (depicted by contours in the lower panels of Fig.~\ref{context}). This ROI corresponds to `ROI 2' investigated by \citet{murray11}, but is not completely identical due to the data preprocessing (i.e., effective smoothing). This paper does not investigate `ROI 1' of \citeauthor{murray11} as it was deemed too small and fragmented to observe any statistically meaningful variations in the extrapolated field. Extrapolation solutions were calculated for each of the four scans in computational volumes comprising of $455 \times 455 \times 228$~pixels$^3$ (i.e., $105\times105\times52$~Mm$^3$). It should be noted that the zoomed-in region (as shown in Fig.~\ref{context}) for further consideration in Sect.~\ref{energy_res} makes use of the full height of the computational volume, covering $165\times135\times228$~pixels$^3$ (i.e., $38\times31\times52$~Mm$^3$). Also note that the values of LFF $\alpha$ obtained for the zoomed-in FOV of each scan are $[0.23, 0.26, 0.22, 0.17]~\mathrm{Mm}^{-1}$ for scans 1, 2, 3, and 4, respectively.


\section{Results}
\label{results}

\begin{figure*}[!]
\centering
\includegraphics[width=\textwidth]{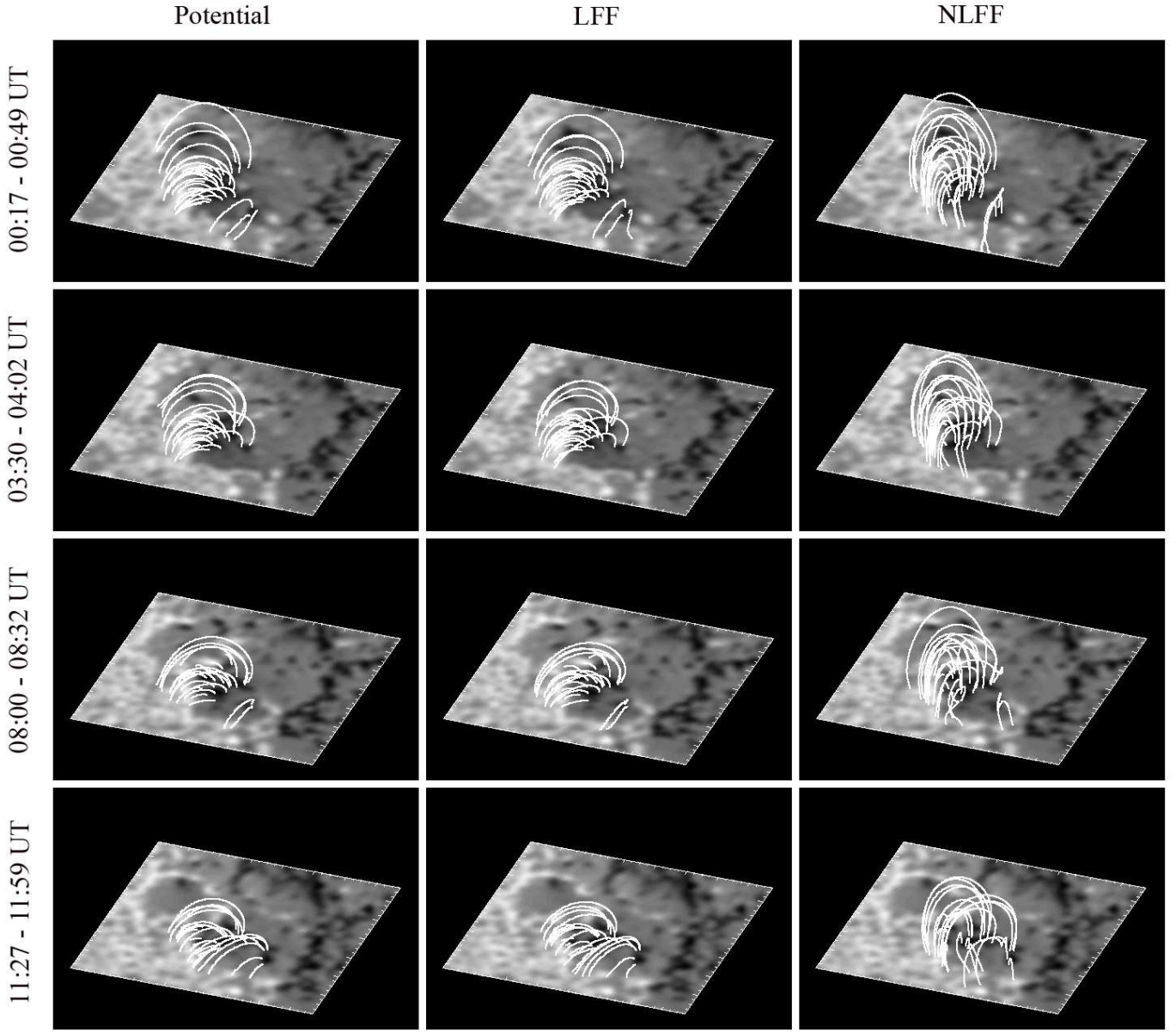}
\caption{Zoomed-in FOV of extrapolated volume (see Fig.~\ref{context}) at increasing scan time from left to right. Upper to lower row: potential field extrapolation, LFF extrapolation, and NLFF extrapolation. Note that the FOV is the same as the $\sim 50" \times 40"$ box shown in Fig.~\ref{context}.}
\label{extrap}
\end{figure*}

Figure~\ref{extrap} shows zoom-ins on the results of the three types of extrapolation from the full FOV, with increasing scan time from left to right (the flare occurs between scans 3 and 4). Field-line traces for every 30$^\mathrm{th}$ pixel in the ROI identified in the bottom panel of Fig.~\ref{context} are shown. The potential and LFF field-line solutions look similar, while the NLFF field lines reach greater heights and connect further South.

In order to study relaxation of the coronal field, it is necessary to determine differences in the field structure between the three forms of extrapolation solution. 
As noted in Sect.~\ref{obs}, the NLFF code used here does not compute $\alpha$. An estimate of $\alpha$ for all pixels within the computational volume can be obtained using Equation~\ref{nlff} ($\nabla \times B = \alpha B$) with the results of the NLFF extrapolation procedure.

\begin{figure}
\centering
\includegraphics[width=\columnwidth]{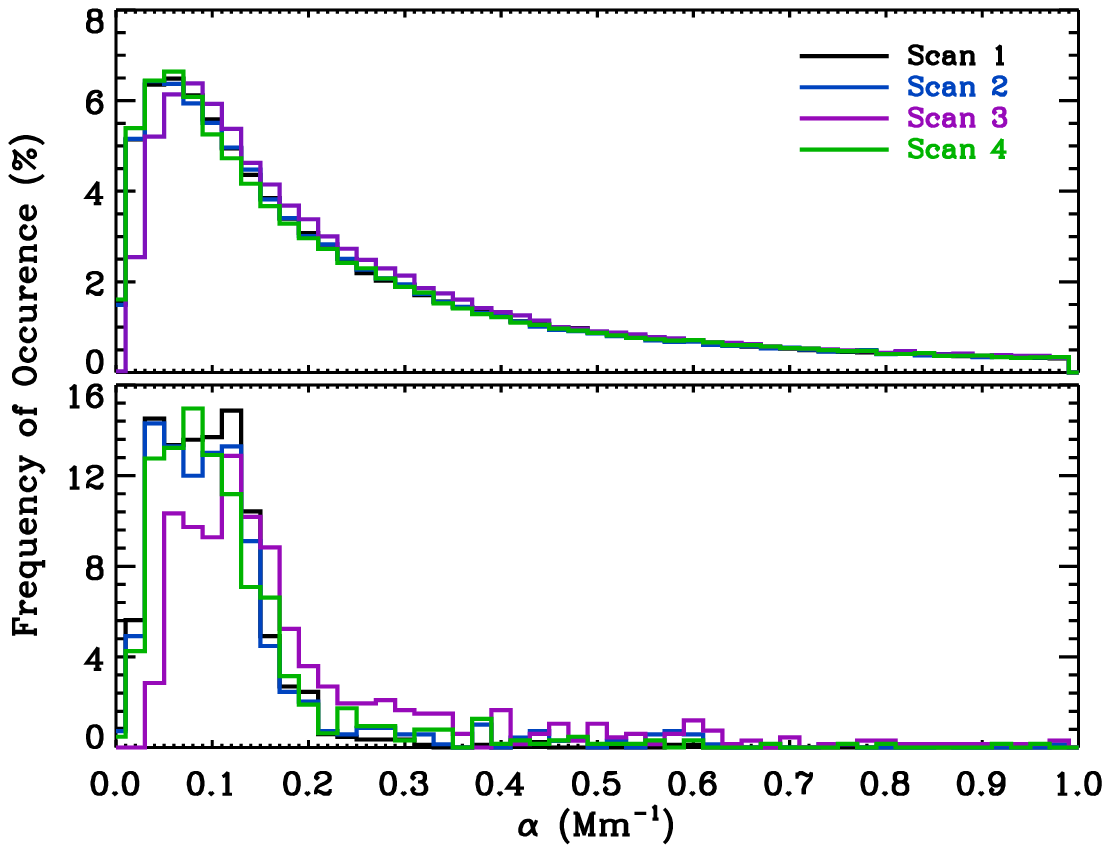}
\caption{Distribution of calculated NLFF $\alpha$ values from all pixels within the full computational volume (upper) and ROI (lower). A bin size of 0.02~Mm$^{-1}$ was used. Ordinate show percentage occurrences, while the abscissa axis indicates the value of $\alpha$ in Mm$^{-1}$. Each scan distribution is coloured as per the legend.}
\label{3D:alpha_hists}
\end{figure}

Average values for $\alpha$ were calculated for each traced field line within the full computational volume as well as in the ROI, and histograms of the results are shown in Figure~\ref{3D:alpha_hists}. For NLFF extrapolations $\alpha$ is assumed constant along the entire field line, and this allows examining the traced field line averages from the source pixels within a region. A consistency check was carried out on full field lines traced from the source pixels, and $\alpha$ was found to be constant within $\pm~0.02$~Mm$^{-1}$. Figure~\ref{3D:alpha_hists} shows $\alpha$ to be very similar for all scans in the full computational volume, with the distributions peaking between $\sim 0.03 - 0.09$~Mm$^{-1}$, and a slight variation in scan 3 towards larger values. For the ROI, $\alpha$ also seems to be similar between scans but a larger variation is observed, and distributions peaking between $\sim 0.03 - 0.13$~Mm$^{-1}$. The distribution for the third pre-flare scan differs the most from the other scans, being shifted towards larger values of $\alpha$. This indicates an increase in the amount of twist in the field in the hours leading up to the flare. The fourth scan distribution indicates a decrease in $\alpha$, and hence twist, after the flare has occurred.

Although the changes in $\alpha$ observed in the full and ROI FOVs over the four scans are not large, it is promising to find expected increases (decreases) before (after) the flare. A number of other 3D magnetic field vector parameters will be examined here, showing more significant changes over the observation period. In particular, geometrical differences will be investigated between field-line traces that are caused by the differing presence of currents in each extrapolation solution. Section~\ref{loops} presents an analysis of 3D spatial offsets between the extrapolation solution field lines, while differences in field-line footpoint locations are presented in Sect.~\ref{res1}. Finally, the total magnetic and possible free magnetic energies are calculated in Sect.~\ref{energy_res}.

\subsection{Spatial Differences in Field-line Traces} 
\label{loops}
To obtain an indirect indication of how $\alpha$ may be varying in the NLFF extrapolation solution, a first step is to determine to what degree the extrapolation solutions differ when considering field-line traces from the same starting pixel. This can be quantified in terms of offsets in the computational $x$, $y$, and $z$ spatial coordinates at specific points along the length of a field-line trace. It should be noted that field lines traced from the same source pixel (i.e., from the ROI) may have different path lengths in each extrapolation solution through 3D space before they return to the photospheric boundary. In order to calculate 3D displacements between comparable locations along each field line, the variable length arrays of ($x$, $y$, $z$) coordinates for each field line trace were interpolated to 11 points. These correspond to relative locations separated by 1/10 of the total path length. Displacements were calculated in each of the $x$, $y$, and $z$ spatial coordinates between the same relative locations (i.e., 1/10 along the potential trace minus 1/10 along the LFF and NLFF traces, etc\ldots) with the values along one entire field-line trace being averaged. The larger this displacement value is the further apart the LFF and NLFF traces are from the potential case, and the more current there is in the system.

\begin{figure*}[!t]
\centering
\includegraphics[width=0.95\textwidth]{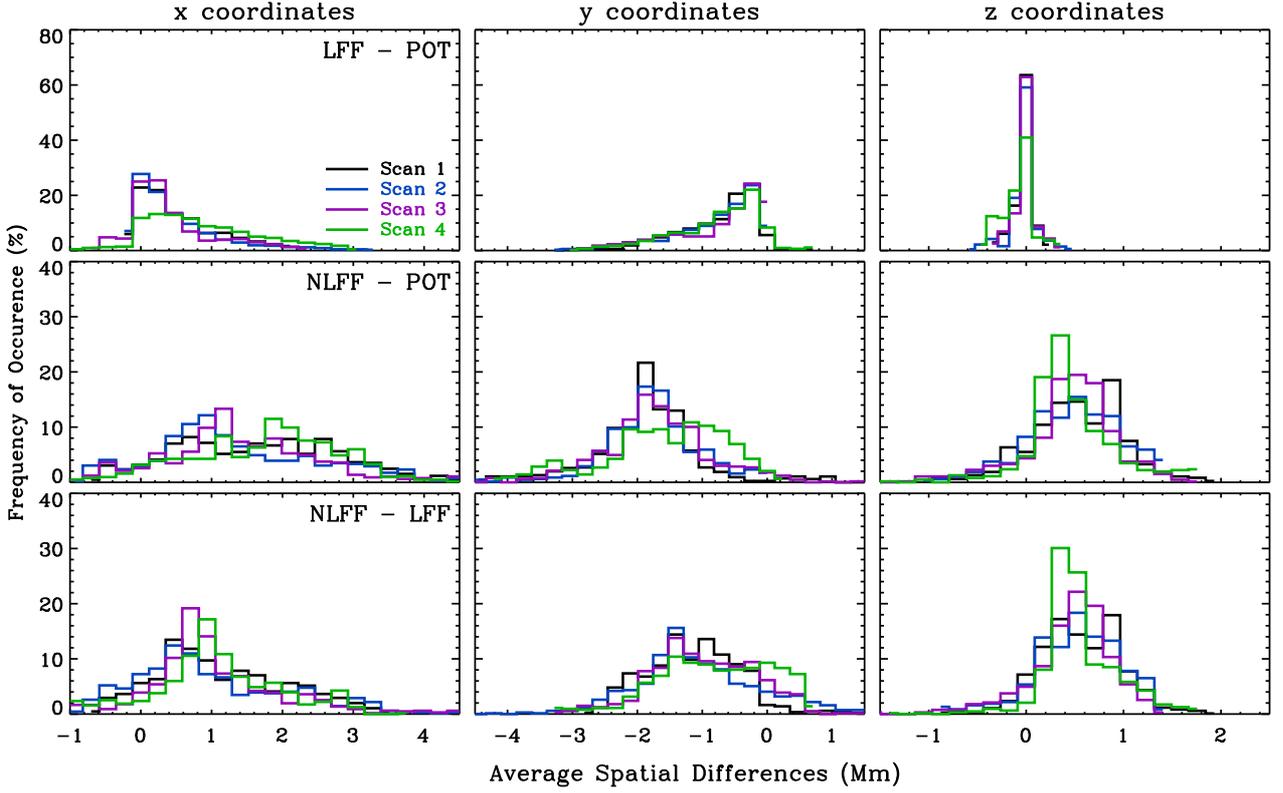}
\caption{Differences in $x$ (left), $y$ (centre) and $z$ (right) coordinates of field-line traces between LFF and potential (upper), NLFF and potential (middle), and NLFF and LFF extrapolations (lower). Each scan distribution is coloured as per the legend, while bin sizes of 1.0 were used for all $x$ and $y$ coordinates, 0.5 for (LFF $-$ POT) $z$ coordinates, and 0.75 for (NLFF $-$ POT) and (NLFF $-$ LFF) $z$ coordinates. }
\label{interp}
\end{figure*}

Figure~\ref{interp} shows the results of differencing and averaging these arrays of $x$, $y$, and $z$ interpolated coordinates for each ROI originating field-line trace. Histogram values for all four scans are overlaid for comparison. Note that all histograms throughout the paper have been normalised to the total number of pixels of the ROI, so as to show the percentage number of pixels. In terms of $x$ coordinates, the mainly positive distributions in the upper left panel show the potential traces generally reach further East than the LFF traces (i.e, reaching smaller $x$ pixel values). The distributions for the (NLFF -- POT) and (NLFF -- LFF) cases are more broad (due to the NLFF solution effectively having a distribution of $\alpha$ values), with the NLFF field lines not reaching as far East as the LFF field lines. In terms of $y$ coordinates, the LFF field generally reaches further South than the potential configuration (shown by the mainly negative $y$-coordinate distributions in the upper middle panel). The central and lower-middle panels again show broader distributions, with the NLFF field-line traces reaching much further South than the potential and LFF traces.

\begin{figure}[!]
\centering
\includegraphics[width=\columnwidth]{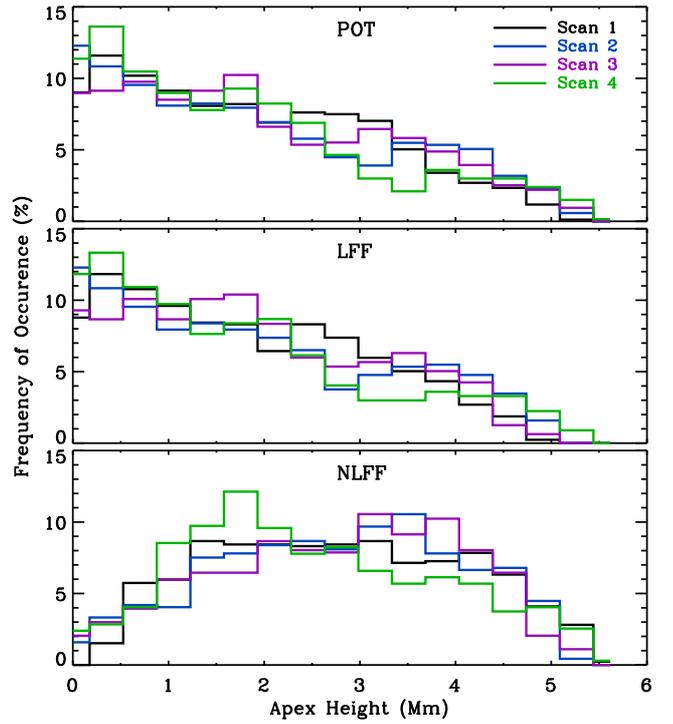}
\caption{Distribution of apex heights from all ROI pixels for potential (upper), LFF (middle), and NLFF (lower) extrapolations. Y-axes show percentage occurrences, while the x-axis indicates the height in Mm. Each scan distribution is coloured as per the legend.}
\label{apex_height}
\end{figure}

\begin{figure}[!t]
\centering
\includegraphics[width=\columnwidth]{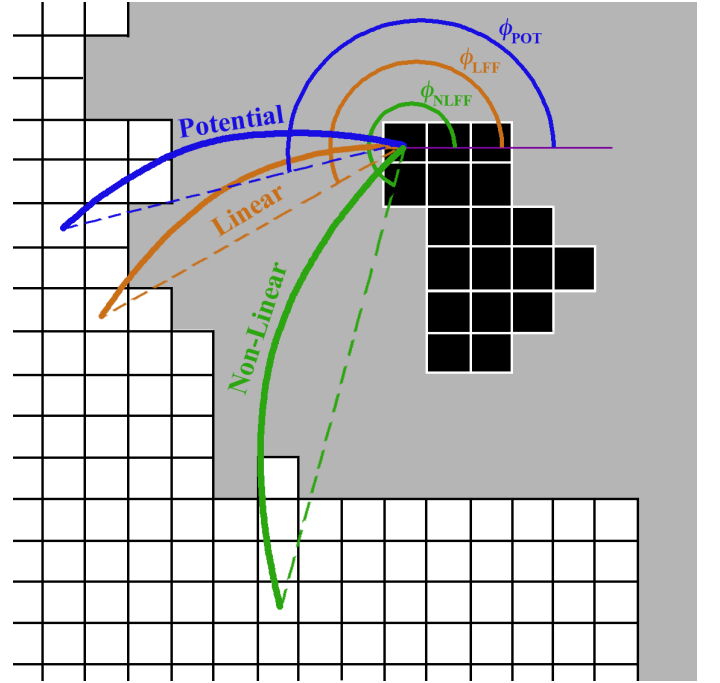}
\caption{Cartoon of typical extrapolated field line traces. The black area indicates the negative polarity ROI, and white area the plage region. The grids outline single pixels. Field-line traces and position angles are indicated in blue (potential), orange (LFF), and green (NLFF).}
\label{cartoon}
\end{figure}

The $z$-coordinate distributions for (LFF -- POT) are much narrower than those of the (NLFF -- LFF) and (NLFF -- POT) distributions, with a greater percentage of values close to zero. The height differences between the NLFF and both the potential and LFF field-line traces indicate that the NLFF field lines exist at larger heights on average within the computational volume. The z-coordinate heights were investigated further, by calculating the maximum field-line height for all pixels within the ROI (see Fig.~\ref{apex_height}). No significant changes are observed with time for the potential or LFF extrapolations. The largest changes are observed in the NLFF results, with the apex height increasing before the flare in scan 3, and then decreasing afterwards in scan 4. Note that NLFF fields reaching greater heights has been previously found \citep{regnierpriest07aa, liu11}. The characteristics derived from Fig.~\ref{interp} are summarised in Fig.~\ref{cartoon}, which illustrates typical field-line traces from an ROI pixel where the NLFF field has a larger magnitude of $\alpha$ (and hence a greater degree of twist) than that of the LFF field.

Considering the variation of Fig.~\ref{interp} across time, none of the coordinate difference distributions vary significantly from scans 1 to 3, whereas larger changes are observed in scan 4 (i.e., after the flare). This is clearest for the $y$ coordinates when considering the central and lower-middle panels. A portion of each of these distributions have shifted towards smaller values, with a greater percentage of the (NLFF -- LFF) distribution having values of zero compared to the (NLFF -- POT) distribution. Thus, after the flare some of the NLFF field lines lie closer in 3D space to the LFF field lines than to the potential ones. This confirms that the LFF field contains current (as it deviates from the current-free potential field) but that the NLFF field has stronger currents (as it deviates further). Smaller differences between the NLFF and LFF fields in the post-flare scan indicates a decrease in NLFF currents relative to those in the LFF field. However, some of this is due to the increased LFF $\alpha$ in scan 4 (Sect.~\ref{obs}). Even though portions of the extrapolation solutions are similar after the flare, the NLFF field contains stronger currents than the LFF field.

\subsection{Orientation Differences Between Field-line Footpoints}
\label{res1}

As indicated in the previous subsection, the amount of current present in the coronal magnetic field affects the direction of field-line traces and ultimately the location of their footpoints (resulting from the twist that currents introduce in the magnetic field). In order to get a better picture of the effective distribution of $\alpha$ values in the NLFF solutions, the footpoint position angle, $\phi$, of a field-line trace is defined as,
\begin{equation}
\phi= tan^{-1}\left(\frac{y_{\mathrm{f}}-y_{\mathrm{i}}}{x_{\mathrm{f}}-x_{\mathrm{i}}} \right) \ ,
\end{equation}
where $x$ and $y$ are pixel coordinates on the solar surface, subscript `i' denotes initial coordinate, and subscript `f' denotes final coordinate. The initial coordinates are taken as each pixel within the negative polarity ROI, such that the footpoint position angle measures the counter-clockwise angle from Solar West to the traced field-line end footpoint in the positive polarity plage (see Fig.~\ref{cartoon} for an illustration).

\begin{figure}
\centering
\includegraphics[width=\columnwidth]{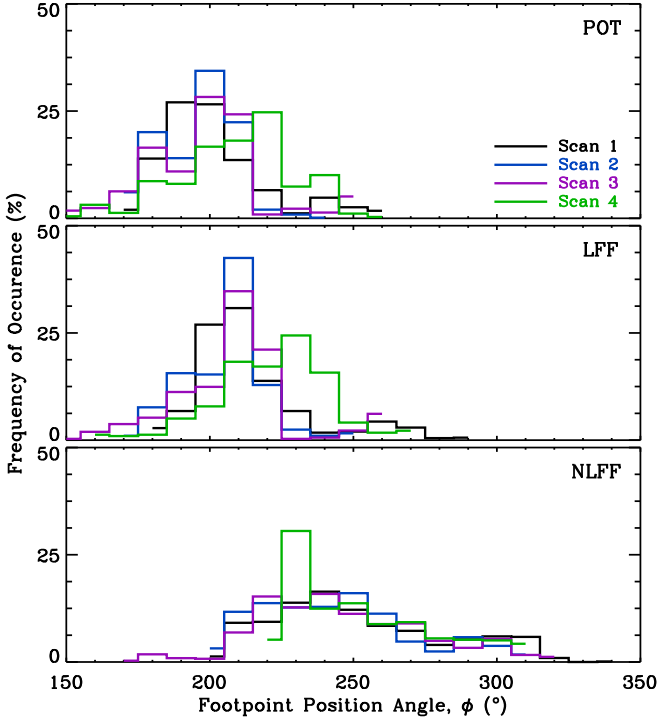}
\caption{Distribution of footpoint position angles from all ROI pixels for potential (upper), LFF (middle), and NLFF (lower) extrapolations. Y-axes show percentage occurrences, while the x-axis gives the end footpoint position angle in degrees counter-clockwise from Solar West. Each scan distribution is coloured as per the legend, while bin sizes of 10$^\circ$ were used throughout.}
\label{azimuth}
\end{figure}

Figure~\ref{azimuth} contains the results of this calculation for each of the extrapolation formats from all four scans. The position angle distribution for the potential case ($\phi^\mathrm{POT}$; upper panel) does not vary much over the first three scans, but a greater portion of the distribution occurs at larger angles after the flare. The position angle distribution for the LFF case ($\phi^\mathrm{LFF}$; middle panel) is very similar to $\phi^\mathrm{POT}$, but shifted to larger angles by $\sim$10--15$^{\circ}$ in all four scans. This is expected with the introduction of twist in the field from the presence of currents that are equally distributed throughout space for the LFF case. Finally, the position angle distributions for the NLFF case ($\phi^\mathrm{NLFF}$; lower panel) are shifted further and broadened. This is also expected, with the NLFF case having a distribution of effective $\alpha$ values throughout space (and hence a distribution of currents and twist in the field). The three pre-flare scans exhibit very similar $\phi^\mathrm{NLFF}$ distributions. However, in scan 4 after the flare the $\phi^\mathrm{NLFF}$ distribution has decreased occurrence of lower angle values and shows a stronger peak at a similar position angle to the peak of the $\phi^\mathrm{LFF}$ distribution. It should be noted that significant portions of the $\phi^\mathrm{LFF}$ and $\phi^\mathrm{NLFF}$ distributions do not overlap.

\begin{figure*}
\centering
\includegraphics[width=0.85\textwidth]{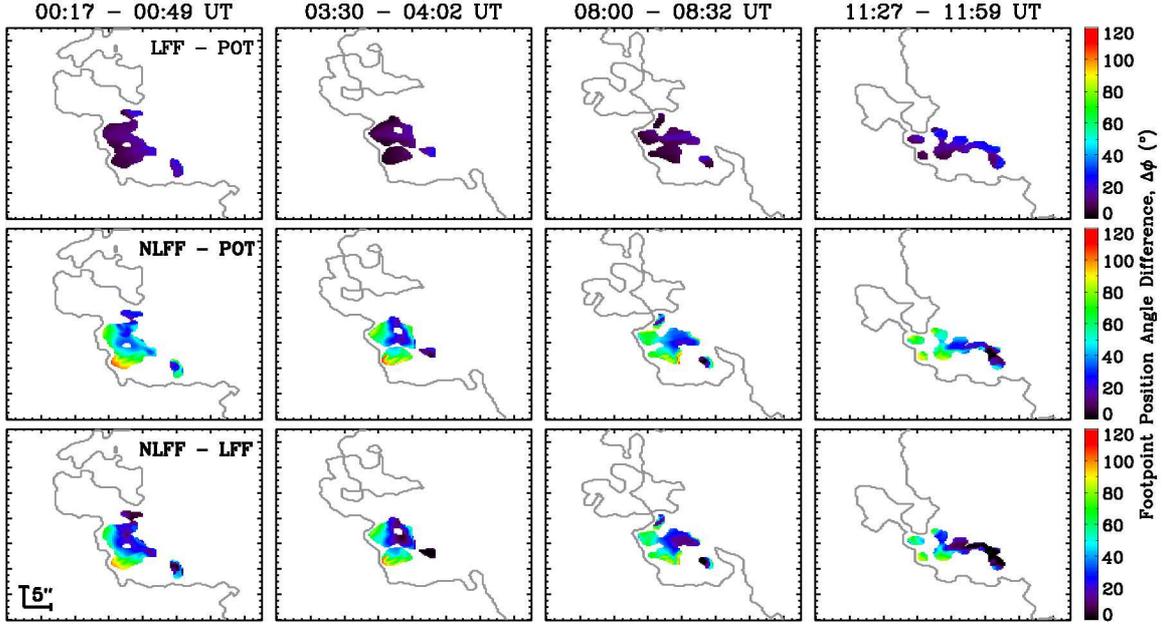}
\caption{Spatial variation of footpoint position angle difference, $\Delta\phi$, from originating ROI pixel locations (see Fig.~\ref{hist} for value distributions). Scan time increases from left to right, while rows depict values of $\Delta\phi^{\mathrm{LFF - POT}}$ (upper), $\Delta\phi^{\mathrm{NLFF - POT}}$ (middle), and $\Delta\phi^{\mathrm{NLFF - LFF}}$ (lower). A neutral line is plotted in grey for context, as seen in the lower row of Figure~\ref{context}.}
\label{mask}
\end{figure*}

The footpoint orientations and temporal variation reported so far correspond to non-localised characteristics of the magnetic field. The spatial distribution of differences between footpoint orientations from the different extrapolation solutions contains information on what portions of the ROI harbour strong currents (and hence magnetic energy). For each ROI source pixel, footpoint position angle values for the three extrapolation traced field-line solutions were subtracted from one another to give the footpoint position angle difference, $\Delta\phi$. This calculated quantity is presented in Fig.~\ref{mask}, where larger values correspond to greater deviation in footpoint position angle between the selected pair of extrapolation solutions (i.e., greater twist in the field). The $\Delta\phi^\mathrm{LFF - POT}$ values are generally small over all scans, with a slight increase in scan 4. This again confirms the known increase of the LFF constant $\alpha$ value in that scan (Sect.~\ref{obs}). Larger quantities are observed for $\Delta\phi^\mathrm{NLFF- POT}$ and $\Delta\phi^\mathrm{NLFF- LFF}$, with the largest located in the South East of the ROI (i.e., nearest to the magnetic neutral line with the positive plage). This indicates that the portion of the ROI containing the strongest currents (thus magnetic energy) is that closest to the neutral line involved in the flare.

\begin{figure*}
\centering
\includegraphics[width=0.85\textwidth]{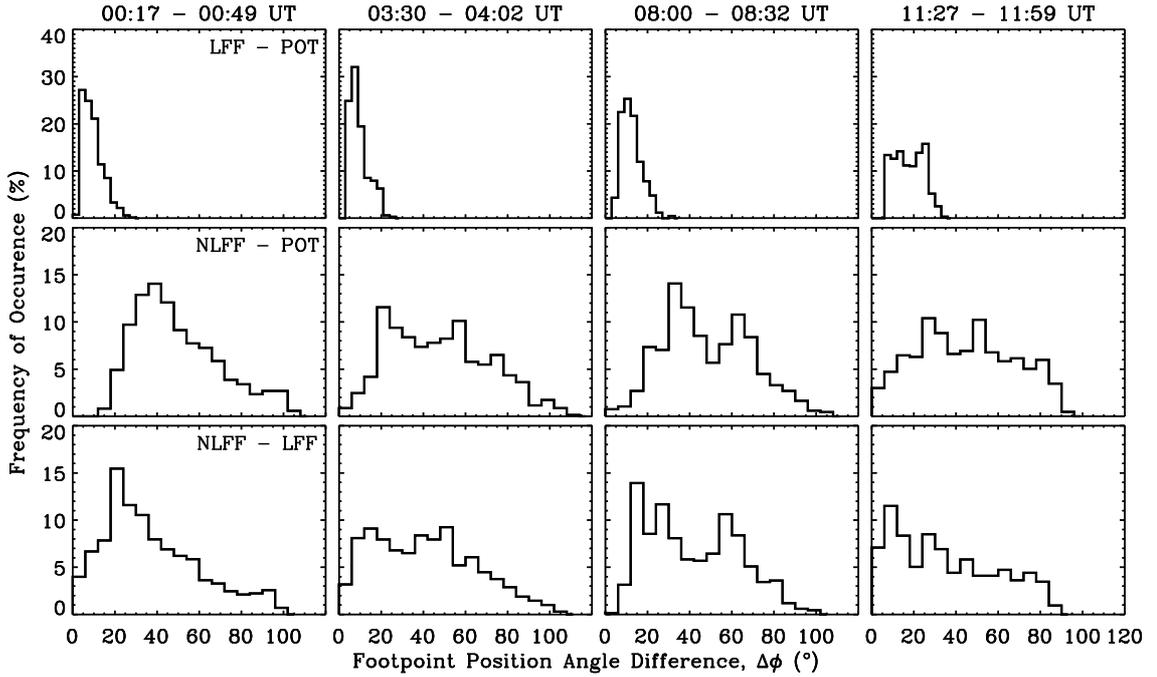}
\caption{Distributions of footpoint position angle difference, $\Delta\phi$ (see Fig.~\ref{mask} for spatial representation). Scan time increases from left to right, while rows depict values of $\Delta\phi^{\mathrm{LFF - POT}}$ (upper; bin size 3$^\circ$), $\Delta\phi^{\mathrm{NLFF - POT}}$ (middle; bin size 6$^\circ$), and $\Delta\phi^{\mathrm{NLFF - LFF}}$ (lower; bin size 6$^\circ$).}
\label{hist}
\end{figure*}

Figure~\ref{hist} represents the footpoint position angle differences as histograms to aid in a more quantitative comparison. The upper row clearly shows the fairly constant, narrow distribution of $\Delta\phi^\mathrm{LFF - POT}$ over scans 1 to 3, which broadens and shifts to larger difference values after the flare (as indicated in Fig.~\ref{mask}). Both the $\Delta\phi^\mathrm{NLFF - POT}$ (middle row) and $\Delta\phi^\mathrm{NLFF - LFF}$ (bottom row) distributions are considerably broader, showing some minor variation over the three pre-flare scans. However, both distributions have a greater percentage of ROI pixels with angle differences close to zero in scan 4 after the flare. The offset observed between the LFF and potential cases in Fig.~\ref{azimuth} results in the $\Delta\phi^\mathrm{NLFF - LFF}$ distribution being shifted to lower values by $\sim$10--15$^\circ$ from the $\Delta\phi^\mathrm{NLFF - POT}$ distribution in all scans. Combining both of these results, the NLFF and LFF post-flare fields appear to be closer in configuration than the NLFF and LFF pre-flare fields were.
 

\subsection{Magnetic and Free Magnetic Energies}
\label{energy_res}

It is important to investigate the amount of energy stored in a magnetic configuration, as some of this stored energy is released in driving the flaring process. The analysis up to now has been concerned with studying geometrical differences between field-line traces in the three extrapolation solutions. However, the effects that the potential, LFF, and NLFF values of $\alpha$ have on the coronal energy content can be calculated directly from the extrapolation solutions. The magnetic energy, $E_\mathrm{m}$, contained in the field is \citep{schmidt64},
\begin{equation}
\label{mag_en}
E_\mathrm{m}=\int_{V}\frac{B^{2}}{8\pi}dV \ ,
\end{equation}
where $V$ is the 3D computational volume under consideration and $B$ is the magnitude of the magnetic field vector at every point in computational ($x$, $y$, $z$) space. However, the values of $E_\mathrm{m}$ for each of the extrapolation solutions does not correspond to the total energy available for flaring \citep{aly84}. Instead, two forms of free magnetic energy, $\Delta E_{\mathrm{m}}$, can be calculated, 
\begin{equation}
\label{nlff-pot}
\Delta E_{\mathrm{m}}^{\mathrm{NLFF-POT}}=E_\mathrm{m}^{\mathrm{NLFF}}-E_\mathrm{m}^{\mathrm{POT}} \ ,
\end{equation}
\begin{equation}
\label{nlff-lff}
\Delta E_\mathrm{m}^{\mathrm{NLFF-LFF}}=E_\mathrm{m}^{\mathrm{NLFF}}-E_\mathrm{m}^{\mathrm{LFF}} \ ,
\end{equation}
which correspond to using either the potential (Eqn.~\ref{nlff-pot}) or the LFF (Eqn.~\ref{nlff-lff}) field configuration as the minimum energy state. Note that there is no free magnetic energy in a potential field configuration, but the LFF and NLFF configurations have free energy due to the presence of currents. Values for $\Delta E_\mathrm{m}^{\mathrm{LFF-POT}}$ are not computed here since the ultimate aim of this work is to study the relaxation of the NLFF field over the course of a flare.

\begin{figure*}
\centering{\includegraphics[width=\textwidth]{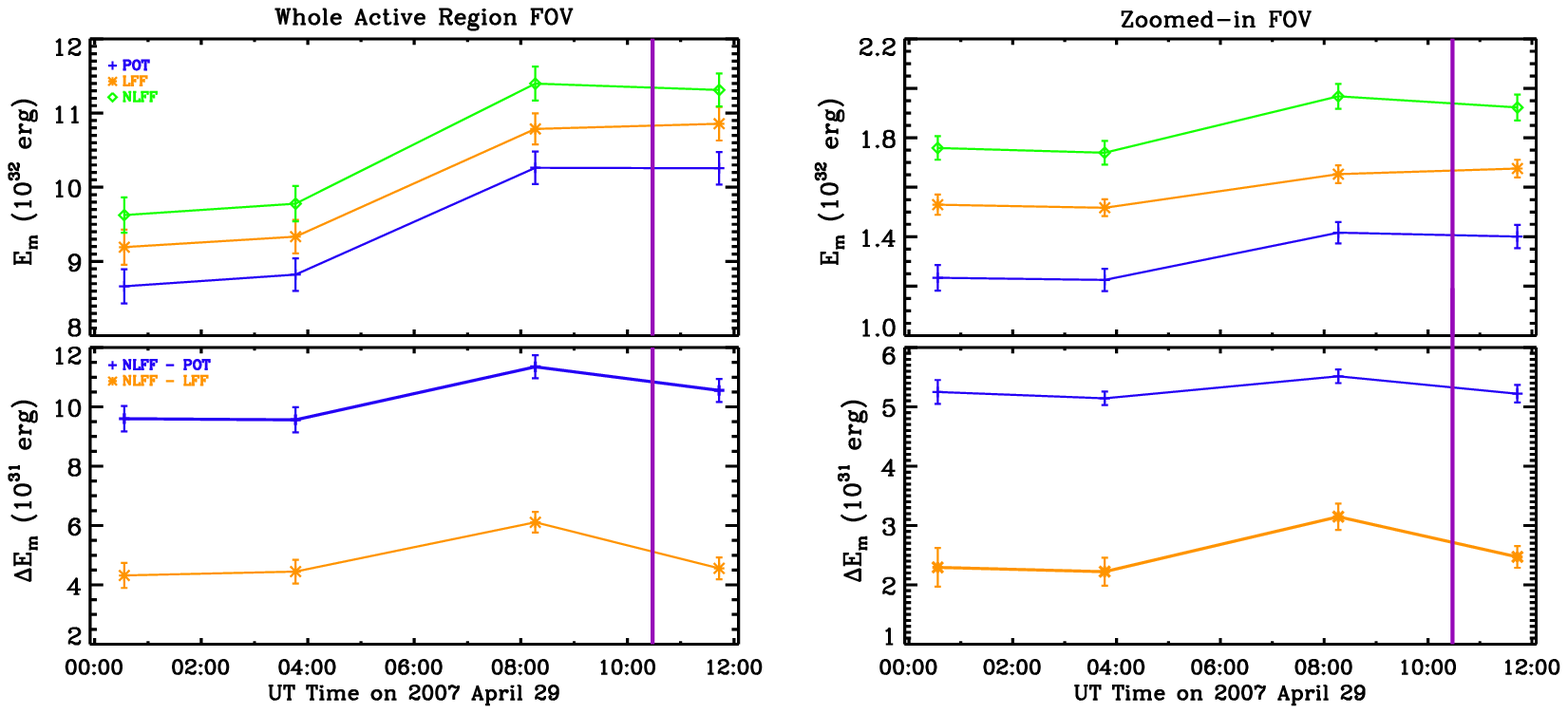}}
\caption{Magnetic energy ($E_\mathrm{m}$, upper) and free magnetic energy ($\Delta E_\mathrm{m}$, lower) for each extrapolation type, as identified in the legends. Energies for the whole active region volume are shown on the left with those for the smaller zoomed-in volume shown on the right, while vertical bars denote errors. The purple vertical lines between the scans 3 and 4 mark the flare peak time.}
\label{energy}
\end{figure*}

Two different volumes are used for the energy calculations: the whole active region volume ($455\times455\times228$~pixels$^3$), and the surface zoom-in represented by the boxes in Figs.~\ref{context} and \ref{extrap} (i.e., $155\times135\times228$~pixels$^3$). Note that the zoomed-in volume uses the whole height of the extrapolation volume. The zoomed-in box was chosen as it corresponds to the location of greatest flare emission in Ca\,\textsc{ii}\,H. The box extent was determined by first choosing the furthest footpoint coordinates of all ROI field-line traces from the four scans as boundary points, but a number of small magnetic flux elements enter and leave this region over the course of the observations. In order to avoid any spurious effects these transient flux elements might have on the calculated energy evolution, the box was enlarged by $\sim$15~pixels on each side. This results in minimal amounts of flux entering or leaving the box over the four scans (Fig.~\ref{context}, lower panels). Note, the zoomed-in region excludes the entire sunspot umbra and penumbra.

The temporal variation of the calculated $E_\mathrm{m}$ values are displayed in the upper panels of Fig.~\ref{energy}, with results from the whole active region volume presented in the left column and the zoomed-in volume in the right column. Note that error bars in this Fig. were determined using estimated uncertainties of $\pm~10$~G for the LOS field and $\pm~100$~G for the transverse field on the photospheric boundary \citep{wiegelmann12}. The percentage errors on the photospheric boundary were assumed to extend upwards, and percentage errors in $B_\mathrm{x}$, $B_\mathrm{y}$, and $B_\mathrm{z}$, for all 3D space were combined. Understandably, values of $E_\mathrm{m}$ from the whole active region volume are much greater than those from the zoomed-in volume. In addition, $E_\mathrm{m}^{\mathrm{NLFF}}$ is consistently larger than $E_\mathrm{m}^{\mathrm{LFF}}$, which is consistently larger than $E_\mathrm{m}^{\mathrm{POT}}$. This corroborates the findings of Sects.~\ref{loops} and \ref{res1}, such that for all scans the NLFF extrapolations contain more twist in the field (i.e., stronger currents and greater magnetic energies) than the LFF extrapolations. For both volumes under consideration, $E_\mathrm{m}$ varies little between the first two scans, but increases in magnitude by scan 3 (i.e., prior to the flare). After the flare, most of the magnetic energies decrease marginally. However, none return to the earlier pre-flare `quiet' values. Differences in $E_{m}^{\mathrm{NLFF}}$ between scans 3 and 4 contain changes in total magnetic energy due to the combined effect of the flare and evolution of the region over a $\sim$3--4~hour period. Between these scans the whole active region volume shows a decrease in $E_{m}^{\mathrm{NLFF}}$ of $(9\pm2) \times 10^{30}$~erg, with the zoomed-in volume decreasing by $(5\pm1) \times 10^{30}$~erg. Both of these changes are likely to be dominated by the flare, given the essentially constant level of total potential magnetic energy observed over the flare.

The results of the free magnetic energy calculations are presented in the lower panels of  Fig.~\ref{energy}. Similar to the total magnetic energies, no significant changes are observed for either volume between the first two scans. Both $\Delta E_\mathrm{m}^{\mathrm{NLFF-POT}}$ and $\Delta E_\mathrm{m}^{\mathrm{NLFF-LFF}}$ increase from scan 2 to 3 (i.e., just prior to the flare). After the flare, $\Delta E_\mathrm{m}^{\mathrm{NLFF-LFF}}$ decreases towards pre-flare `quiet' values, remaining slightly higher than the level observed in scans 1 and 2. $\Delta E_\mathrm{m}^{\mathrm{NLFF-POT}}$ also decreases, but not as much as $\Delta E_\mathrm{m}^{\mathrm{NLFF-LFF}}$. In terms of absolute changes over the flare, $\Delta E_\mathrm{m}^{\mathrm{NLFF-LFF}}$ decreases by $(1.6\pm0.3)\times10^{31}$~erg in the whole active region volume, while the zoomed-in volume decreases by $(7\pm2)\times10^{30}$~erg. Caution should be taken when interpreting changes in $\Delta E_\mathrm{m}^{\mathrm{NLFF-LFF}}$ over time (e.g., it is known here from Sect.~\ref{obs} that the value of LFF $\alpha$ is different before and after the flare). However, it is at least encouraging that the changes in free magnetic energy above the LFF state are within a factor of 2 of the absolute changes in total NLFF magnetic energy. This indicates that $\sim$15-25\% of the free magnetic energy that existed above the LFF state before the flare has been lost from the system over the course of the flare.


\section{Discussion and Conclusions}
\label{concl}

The magnetic field configuration of a ROI within an active region has been studied over the course of a \emph{GOES} B1.0 magnitude flare. Geometrical differences between field-line traces through potential, LFF, and NLFF extrapolation solutions have been analysed, as well as their resulting magnetic energies. It is found that the general orientation of ROI field-line footpoints do not change significantly in the hours leading up to the flare, despite the ROI showing an increase in total magnetic energies. However, there are signatures of field redistribution after the flare that indicate incomplete Taylor relaxation: a portion (i.e., not all) of the NLFF field configuration becomes similar to that of the LFF field. Consideration of the magnetic and free magnetic energies after the flare indicate that the region still has energy available for further flaring. It is worth noting that a \emph{GOES} B1.2 flare occurs in the same spatial region on 2007\,April\,29 at 14:35~UT, less than 3~hours after the final SOT-SP scan analysed here.

A number of previous works have debated the relevance of Taylor relaxation to the flaring process. \citet{amari00} use the presence of non-linearities in the post-relaxation state of numerical simulations to suggest that Taylor's theory does not apply to flares and CMEs. \citet{bleybel02} reach the same conclusion using observations, finding that the relaxed post-eruption state is inconsistent with a LFF state. However, a large amount of helicity was ejected from the region studied. In relation to this, \citet{regnierpriest07apj} state that if helicity is not conserved (e.g., during a CME) then the minimum energy state can be a potential one. It is also not yet known whether helicity conservation is the only constraint on relaxation \citep{pontin11}. Although helicity is not calculated here, it should be noted that the studied flare event has no associated CME.

In contrast, other studies have demonstrated the presence of Taylor relaxation. Numerical simulations of energy release in a coronal loop by \citet{browning08} indicate that the relaxed equilibrium state corresponds closely to a constant $\alpha$ field. \citet{nandy03} report the observational detection of a process akin to partial Taylor relaxation in flare-productive active regions that never achieve completely LFF states within their observation periods. It is worth noting, \citeauthor{nandy03} find that the relaxation process occurs on of the order of a week. The 12~hour period studied here is a considerably shorter time scale, which may contribute to only partial Taylor relaxation being observed. It has also been suggested that partial relaxation is perhaps to be expected from a magnetically complex system with flux emergence and cancellation to be accounted for  \citep{pontin11}.

In the build up to the flare, a clear increase is observed in all magnetic energies and free magnetic energies, occurring  $\sim$6.5--2.5 hours prior to the start time of the event. On average, in the zoomed-in region volume the magnetic energy increases by $\sim 7 \times10^{30}$~erg and free magnetic energy increases by $\sim 2 \times10^{31}$~erg. In the whole active region volume, the magnetic energy increases by $\sim 20 \times10^{30}$~erg and free magnetic energy increases by $\sim 15 \times10^{31}$~erg. This may indicate a shorter time scale for flare energy input than has been previously observed \citep[e.g., a gradual increase in magnetic energy over the course of a day before an M-class flare is reported by][]{thalmannwiegelmann08}, but could be related to the low magnitude of the event studied here. It is worth noting, in studying \emph{Hinode} data associated with X-class flares, \citet{jing10} found no clear and consistent pre-flare pattern in the temporal variation of free magnetic energy above the potential state.

Considering changes over the flare, a marginal decrease is observed in most of the magnetic energies. Previous authors have been primarily concerned with reporting changes in the free magnetic energy above the potential state over solar flares. For example, \citet{sun12} find a decrease in $\Delta E_\mathrm{m}^{\mathrm{NLFF-POT}}$ of $\sim$$3 \times 10^{31}$~erg within 1~hour of an X2.2 flare (which they believe to be an underestimation), \citet{thalmannwiegelmann08} find a decrease of $\sim$$5 \times 10^{32}$~erg over an M6.1 flare, and \citet{thalmannwiegelmann08a} find a decrease of $\sim$$2 \times 10^{31}$~erg after a C1.0 flare ($\sim$40\% of the available free magnetic energy). The general scaling of the flare event magnitudes agree with the B1.0 flare-related changes of (3 or 8)\,$\times10^{30}$~erg in $\Delta E_\mathrm{m}^{\mathrm{NLFF-POT}}$ (for the zoomed-in and whole active region volumes considered here, respectively). However, \citet{regnierpriest07apj} suggest that $\Delta E_\mathrm{m}^{\mathrm{NLFF-POT}}$ gives an upper limit for the energy that can be released during large flares, while $\Delta E_\mathrm{m}^{\mathrm{NLFF-LFF}}$ is a good estimate of the energy available for small flares. In this work the value of $\Delta E_\mathrm{m}^{\mathrm{NLFF-LFF}}$ prior to the flare is $\sim$(3 or 6)\,$\times10^{31}$~erg (for the zoomed-in and whole active region volumes, respectively) with a corresponding decrease over the flare of $\sim$(1 or 2)\,$\times10^{31}$~erg. This indicates that $\sim$20--30\% of the free magnetic energy above the LFF state is removed during the course of the flare.

In summary, it seems that the magnetic configuration of active region NOAA 10953 was did not fully relax to either a potential or LFF state after the B1.0 flare on 2007\,April\,29. In addition, the energy budget remained sufficient to trigger another flare within 4~hours. Finding an active region in a partially relaxed state after previous flaring may then be a good indicator of impending flare activity. Also, the increase of magnetic energy on short time scales before a flare could be useful for near-realtime forecasting. However, the methods used to obtain these quantities are too computationally intensive to be currently applied in near-realtime.

\begin{acknowledgements}
\textit{Hinode} is a Japanese mission developed and launched by ISAS/JAXA, with NAOJ as domestic partner and NASA and STFC (UK) as international partners. It is operated by these agencies in co-operation with ESA and NSC (Norway). S.A.M. is supported by the AXA Research Fund, while D.S.B. is supported by the European Community (FP7) under a Marie Curie Intra-European Fellowship for Career Development. 
\end{acknowledgements}

\bibliographystyle{aa}
 \bibliography{bibliography}  
 
 \end{document}